\documentclass{article}
\usepackage[a4paper,top=1cm,bottom=1cm,left=1.2cm,right=1.2cm]{geometry}

\usepackage{graphicx}% Include figure files
\usepackage{dcolumn}% Align table columns on decimal point
\usepackage{bm}% bold math
\usepackage{amsmath}
\usepackage{lineno}
\usepackage{mathtools}
\usepackage[section]{placeins}
\usepackage{textcomp}
\usepackage{authblk}
\usepackage{xcolor}
\usepackage{braket}
\usepackage{amssymb}
\usepackage{float}
\begin{document}

\title{Non-uniform modal power distribution caused by disorder in multimode fibers}

\author[1*]{Mario Zitelli}

\affil[1]{Department of Information Engineering, Electronics, and Telecommunications, Sapienza University of Rome, Via Eudossiana 18, 00184 Rome, Italy}
\affil[*]{mario.zitelli@uniroma1.it}

\maketitle

 \begin{abstract}
The evolution of modal crosstalk in multimode fibers is investigated using four different experimental and numerical approaches. Results converge to demonstrate that the fiber disorder alone is capable of generating steady states characterized by non-uniform modal power distributions, which promote the lower-order modes at the expenses of the higher-order ones and are properly described by a weighted Bose-Einstein law.
 \end{abstract}

\section{Introduction}

Multimode fibers (MMF) \cite{Gloge:6767859,Hasegawa:80} and multicore fibers (MCF) \cite{Inao:79,Noordegraaf:09} support the simultaneous propagation of orthonormal modes characterized by different field distribution $F_p(x,y)$, propagation constant $\beta_p$ and fiber loss $\alpha_p$. In graded-index multimode fibers (GRIN), the fiber index profile is commonly parabolic-shaped \cite{Agrawal_2023} and the fiber modes compose groups of quasi-degenerate modes with the same propagation constant $\beta_j$.

MMFs and MCFs were rediscovered in an attempt to solve the capacity problem in optical networks \cite{Essiambre:5420239,Winzer:18} using the space-division multiplexing (SDM) technique \cite{Richardson-NatPhot-2013-94-2013}, which adds a further degree of freedom for increasing the transmission capacity by multiplexing channels into the different orthonormal modes or cores of the fiber.

One of the main transmission impairments of MCF and MMF low-power systems arise from fiber imperfections, consisting of micro and macro bending, fiber radius and index random changes, which cause the random-mode coupling effect (RMC) \cite{Gloge:6774107,Savovi2019PowerFI,Ho:14}, characterized by a random exchange of power between orthonormal modes of a fiber. A second detrimental effect is the mode-dependent loss (MDL), where different non-degenerate modal groups experience different linear losses \cite{Gloge:6767859}. As it will be shown, both effects contribute to generating non-uniform modal power distributions, where the lowest-order modal groups (LOMs) are promoted at the expense of the highest-order ones (HOMs).

In the literature, numerical simulations of multimode fibers that include RMC provided conflicting results, depending on the model used for fiber disorder. For example, in  \cite{FERRARO2025134758,PhysRevLett.129.063901} disorder was modeled using a random potential $\delta V(x,y,z)=\mu(z)g(x,y)$, with $\mu(z)$ a stochastic function with zero mean and $<\mu(z)\mu(0)>=\sigma_b^2\exp(-z/L_c)$, being $L_c$ the RMC correlation length and with disorder arising from an elliptical fiber structure of the type $g(x,y)=\cos(b_x x /x_0)\cos(b_y y/y_0)$; this approach provided steady states where the power distribution between modes was nearly uniform.

In \cite{Yadav:10433661} a matrix propagation model was used, where the end-to-end fiber transfer matrix was obtained as the product of several section’s transfer matrix. Different cross-talk patterns were obtained, depending on the model used for the section transfer matrices (homogeneous perturbation drift model, propagation constant drift model).

In this work, we use four different experimental and numerical approaches to study the steady state modal power distribution produced by the RMC in several kilometers of GRIN fiber. Specifically, the four approaches are:

i) Numerical simulations based on coupled generalized nonlinear Schrödinger equations (GNLSE)  \cite{Poletti:08, Wright2018a}, modified to include the effects of multimode fiber disorder. The model adopted for RMC is the well-known field coupling model \cite{Mumtaz_2013,Ho:14}, using modal coupling coefficients depending on fiber index perturbations $\Delta n(x,y,z)$ with uniform random distribution and on overlapping integrals between the transverse fields $F_p(x,y)$ of the coupled modes.

ii) Numerical simulations based on the well-known power-flow equations model (PFE) \cite{Savovi2019PowerFI, Gloge:72, Gloge:6774107}, describing the bi-directional flow of power between fiber modal groups. According to the model, power flows towards LOMs and HOMs are not symmetrical, causing the steady state modal power distribution to be non-uniform.

iii) Experiments were low-power, Gbit/s pulses were transmitted over several kilometers of GRIN fiber and input (output) modes were injected (decomposed) using modal multiplexers based on multi-plane light conversion (MPLC) technology \cite{Labroille201793}. In the experiments, the Kerr and Raman nonlinearities were negligible and RMC was not artificially forced but was caused by the natural imperfections of a commercial GRIN fiber.

iv) Experiments recalling case iii) with low-power pulses replaced by single-photon pulses, and power meters by photon-counting techniques. Those experiments confirmed that a non-uniform output distribution is produced even in the ergodic case of single-photon exchanges between interacting modes.

The four approaches converged providing similar results, confirming that RMC alone is capable of generating steady states characterized by non-uniform modal distributions in a GRIN fiber; the distributions promote the LOMs at the expenses of the HOMs and are properly described by a weighted Bose-Einstein law (wBE) \cite{Zitelli:lpor.19.202400714,Zitelli:10848169}. In each of the four approaches, the MDL was characterized and properly isolated to measure the contribution of the individual effects.

\section{Materials and Methods}

\subsection{Experiments}

\begin{figure}[H]
%\isPreprints{\centering}{} % Only used for preprints
\centering
\includegraphics[width=0.8\textwidth]{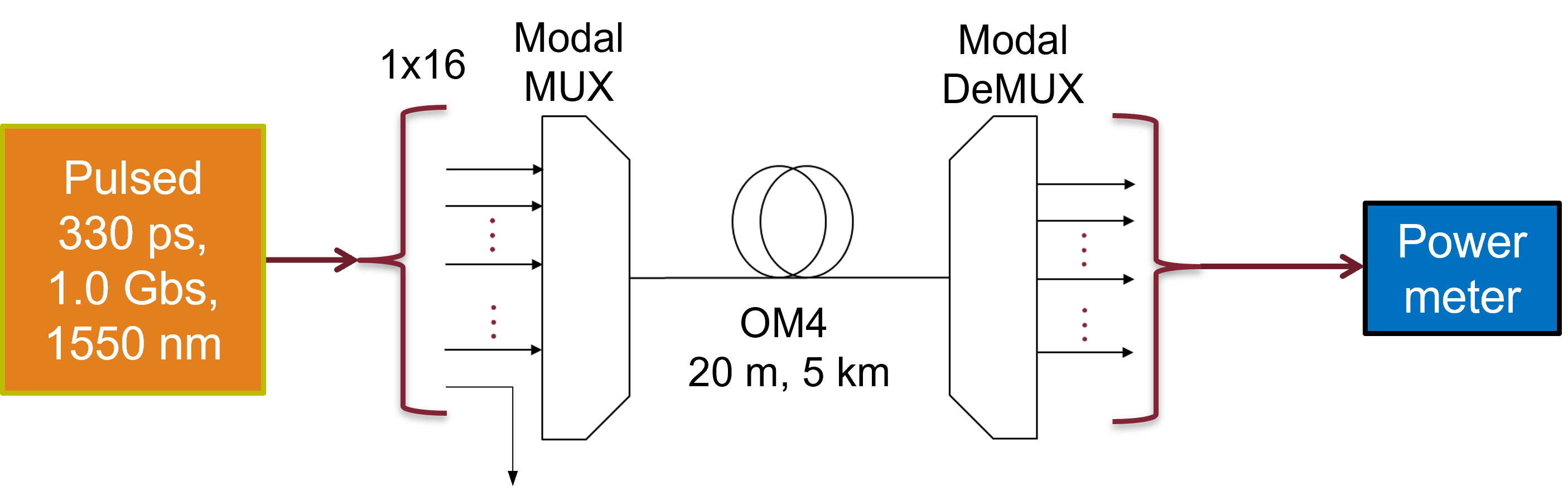}
\caption{The experimental setup for the linear regime.\label{fig1}}
\end{figure}

For the experiment in the classical linear regime, a continuous wave laser at $\lambda=1550$ nm, with a 10 kHz linewidth (Thorlabs TLX1), was carved to a train of pulses at 1 Gb/s by an optical intensity modulator (IM, Exail MX-LN-10) driven by an FPGA, pulsewidth was 330 ps. Signal was split by a 1x16 coupler and applied to different optical delay lines before being coupled to the transmission line (Fig.~\ref{fig1}).

Signals were applied to the different modal inputs of a modal multiplexer (Cailabs Proteus C-15) characterized by multi-plane light conversion (MPLC) technology \cite{Labroille201793}; the multiplexer coupled the power of each input to one of the first 15 Laguerre-Gauss modes ($LG_p$ with $p=1,2,..,15$) of a GRIN OM4 fiber (Thorlabs GIF50E); the fiber was capable of supporting $Q=10$ groups of quasi-degenerate modes at 1550 nm, corresponding to $M=55$ modes per polarization (group $j$ includes $j$ quasi-degenerate modes); hence, only the first 5 groups out of 10 were populated at the input.
One of the input signals was measured by a power meter.

Two spans of OM4 fiber (20 m or 5 km) were alternatively spliced to the multiplexer/demultiplexer. The parameters of the OM4 fiber at the wavelength $\lambda=1550$ nm are: dispersion $\beta_2=-28.0$ (or -28.6) ps$^2$/km for mode 1 (or mode 15), higher-order dispersion $\beta_3=0.15$ ps$^3$/km, material loss $\alpha=5.0\times10^{-5}$ m$^{-1}$, Kerr nonlinear coefficient $n_2=2.6\times10^{-20}$ m$^2$/W, fiber radius $a=25$ $\mu$m. 

At the output, a twin modal demultiplexer was used to decompose the received beam into modes. The power content of the first 15 Laguerre-Gauss modes was measured by a power meter.

The modal power at the input of the OM4 was 7.4 $\mu$W, corresponding to a pulse energy and peak power of 7.4 fJ and 21 $\mu$W, respectively. The dispersion length, where dispersion-induced pulse broadening is not negligible, is $L_D=1402$ km; the nonlinearity length, where the Kerr nonlinearity cannot be neglected, is $L_{NL}=88700$ km; hence, pulses propagated in the linear regime and conserved their shape in the experiment; however, they were subject to material and modal loss and RMC caused power exchange among degenerate and non-degenerate modes. The pulsed regime was used to reduce the phase coherence between different modes, which could cause fluctuations in the decomposed modal power. With this choice, the modal power fluctuation at the output of the demultiplexer was less than 10\%, which represents the accuracy of the experiment.

In the experiment in quantum regime, pulses were attenuated to one photon per pulse per mode at the input of the OM4; input and output power meters were replaced by single-photon detectors (ID Qube NIR) in  gated mode with 10\% efficiency; histograms were counted using a time controller (IDQ ID1000) with 25 ps resolution.

%\begin{quote}
%This is an example of a quote.
%\end{quote}

\subsection{Numerical model for GNLSE}

The generalized nonlinear Schrödinger equation model (GNLSE) for the multimode fiber is a derivation of \cite{Poletti:08, Wright2018a}, modified to include the presence of RMC. The complex amplitude for mode $p$ evolves as

\begin{multline}
\frac{\partial A_p\left(z,t\right)}{\partial z}=i\left(\beta_0^{\left(p\right)}-\beta_0\right)A_p-\left(\beta_1^{\left(p\right)}-\beta_1\right)\frac{\partial A_p}{\partial t}+i\sum_{n=2}^{4}{\frac{\beta_n^{\left(p\right)}}{n!}\left(i\frac{\partial}{\partial t}\right)^nA_p} -\frac{\alpha_p\left(\lambda\right)}{2}A_p
\\ +i\frac{n_2\omega_0}{c} \sum_{l,m,n} Q_{plmn} \{\left(1-f_R\right)A_lA_mA_n^\ast +f_RA_l\left[h\ast\left(A_mA_n^\ast\right)\right]\}    .
\label{eq:GNLSE}
\end{multline}

\noindent The right-hand side terms of Eq. \ref{eq:GNLSE} describe: modal dispersion, four orders of chromatic dispersion, wavelength-dependent losses, nonlinear Kerr and Raman terms, respectively. 
The $Q_{plmn}$ are cross-terms corresponding to the inverse of the effective modal areas, providing appropriate weights to inter-modal four-wave mixing (IM-FWM) and inter-modal stimulated Raman scattering (IM-SRS) terms. 
The Raman term in Eq. \ref{eq:GNLSE} contributes with a fraction $f_R=0.18$; the expression $h\ast\left(A_mA_n^\ast\right)$ denotes time convolution with the Raman response function $h(t)$, with typical time constants of 12.2 and 32 fs \cite{Stolen1989}. In the simulations presented in this work, Kerr and Raman nonlinearities are included although they produce no effect.

Disorder-induced RMC is accounted for by a simulation step introduced every $L_c$ meters, being $L_c$ the RMC correlation length; the applied equation is

\begin{equation}
\frac{\partial A_p\left(z,t\right)}{\partial z}= i\sum_{m \ne p}{q_{mp}A_m\left(z,t\right)}    .
\label{eq:RMCeq}
\end{equation}

The coefficients $q_{mp}$ account for linear RMC, and are associated with random imperfections of the refractive index profile due to macro and micro bending, core deformation and index random changes.  Here we suppose that RMC occurs both among quasi-degenerate and non-degenerate modes; they are calculated at each step as \cite{Mumtaz_2013,Ho:14}

\begin{equation}
q_{mp}=\frac{\pi}{\lambda n^{(p)}_{eff}\sqrt{I_mI_p}}\int\int F_m(x,y)F_p(x,y)\Delta n^2(x,y)dxdy      ;
\label{eq:RMCcoeff}
\end{equation}

in Eq.~\ref{eq:RMCcoeff}, $F_m$ and $F_p$ are the modal transverse field profiles with intensity integrals $I_m$, $I_p$, respectively; $n^{(p)}_{eff}$ is the effective index for mode $p$, and the random fluctuation of core index, which models the disorder, is related to a random variable $rand(x,y)$ uniformly distributed between 0 and 1 as

\begin{equation}
\Delta n(x,y)=[2 rand(x,y)-1]\sqrt{RMC_{ex}}    ,
\label{eq:Deltan}
\end{equation}

being $RMC_{ex}$ a RMC extent for $\Delta n^2(x,y)$. The index distribution $\Delta n(x,y)$ is re-calculated at each step.

After Eq.~\ref{eq:RMCeq} is applied during an RMC step, the total power is controlled so that it does not undergo discontinuity with respect to the value before the step.

\subsection{Numerical model for Power Flow Equations}

In the linear regime, RMC in multimode fibers is properly modeled by the well-known power-flow diffusive equations (PFE); the model does not account for modal degeneracy, and applies to the total power in modal groups $j=1, 2, .., Q$. If $P_j$ is the power of the $j$-th mode group, the power exchange among adjacent non-degenerate modal groups is described by \cite{Savovi2019PowerFI, Gloge:72, Gloge:6774107}

\begin{equation}
P_j(z+L_c)=\big(\frac{DL_c}{\Delta m^2}-\frac{DL_c}{2m\Delta m}\Big)P_{j-1}(z)+\Big(1-\frac{2DL_c}{\Delta m^2}-(\alpha_0+Am^2)L_c\Big)P_j(z)+
\big(\frac{DL_c}{\Delta m^2}+\frac{DL_c}{2m\Delta m}\Big)P_{j+1}(z)   ,
\label{eq:PowerFlow}
\end{equation}

with $m(j)=j-1$, $D$ and $\alpha_0$ (1/m) the coupling coefficient and linear losses, respectively, $A$ the modal loss coefficient, $\Delta m=1$ the modal step and $L_c$ the RMC integration step. In Eq.~\ref{eq:PowerFlow}, power flows in both directions, from group $j$ down to group $j-1$ and up to $j+1$; after consecutive integration steps, a cascading effect is produced causing the coupling of non-adjacent groups. 

Mode coupling among quasi-degenerate modes into groups is neglected, because it is so fast that statistical modal equipartition into groups can be assumed; it is then meaningful to calculate the mean modal content into groups as $\lvert f_j \rvert^2=2P_j/(g_jP_{tot})$, with $g_j=2j$ the group degeneracy and $P_{tot}$ the total power in the multimode fiber.

%%%%%%%%%%%%%%%%%%%%%%%%%%%%%%%%%%%%%%%%%%
\section{Results}

\subsection{Linear Regime Experiment}

As a first test in the linear regime, material loss was measured by coupling power to the fundamental mode $LG_1$ and measuring the total output power from 20 m and 5 km of fiber. The comparison provided $\alpha=0.19$ dB/km.
Insertion loss of modal groups was measured by applying power to all quasi-degenerate modes of a group and measuring the total output power using 20 m of fiber; this included the loss from the multiplexer/demultiplexer devices, while the fiber loss was negligible. The resulting transmittances, in linear scale, were $IL_j=0.49, 0.36, 0.38, 0.38, 0.37$ for groups $j=1,2,..,5$, respectively.

Mode-dependent loss (MDL) was measured by applying power to and measuring power from all modes of a group, using 5 km of fiber, and normalizing to material loss and insertion loss.
The group transmittance related to the modal loss was found to be properly described by $MDL_j=exp[-A(j-1)^2L]$ with $L=5000$ m and $A=4.4\times10^{-6}$ m$^{-1}$.

After having characterized the system losses, equal power was applied to the modes of one group $j$, or to all input modes, and the output power was measured from the individual modes $p$.
Output power $P_p$ was measured from the 15 modes, calculating the mean and standard deviations over 60 s acquisitions; the output mean modal power fraction was calculated dividing the insertion loss $IL_j$ and the modal loss $MDL_j$ and respecting the condition $\sum_{j=1}^5 j\lvert f_j \rvert^2=1$ 

\begin{equation}
\lvert f_j \rvert^2=\frac{1}{j IL_j MDL_j P_{tot}} \sum_{p=j(j+1)/2-j+1}^{j(j+1)/2}P_p .
\label{eq:MMPF}
\end{equation}

This was done with the purpose to estimate the power transfer between modal groups caused solely by RMC, by normalizing modal and insertion losses.

\begin{figure}[H]
%\isPreprints{\centering}{} % Only used for preprints
\centering
\includegraphics[width=12.0 cm]{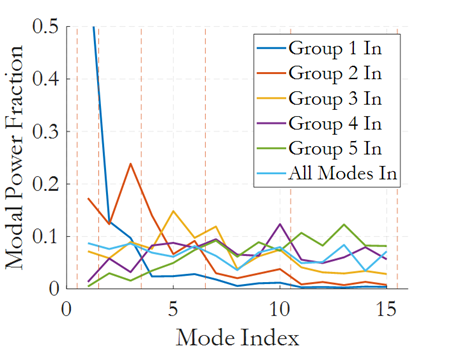}
\caption{Experimental output fractional modal power normalized to insertion and modal loss, when individual groups or all input modes are excited.\label{fig2}}
\end{figure}

\begin{figure}[H]
%\isPreprints{\centering}{} % Only used for preprints
\centering
\includegraphics[width=12.0 cm]{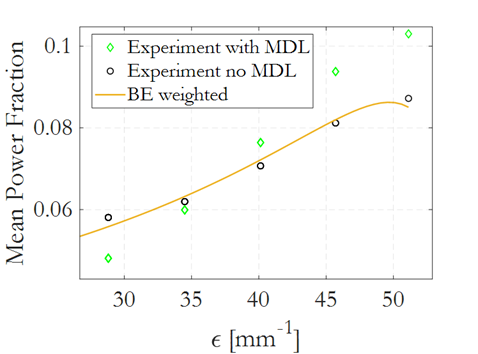}
\caption{Experimental mean modal power fraction of the output groups after uniform excitation of the input modes. A wBE fit is also shown.\label{fig3}}
\end{figure}

Figure~\ref{fig2} provides the modal fractional power after normalizing to $IL_jMDL_j$, when power is applied to the individual modal groups or to all modes. Curve 'All Modes In' in particular shows a non-uniform power distribution at the output; LOMs are promoted by an asymmetric power flow between modal groups, which will be discussed later. This property is more evident in Fig.~\ref{fig3}, where the average modal power fractions are calculated using Eq.~\ref{eq:MMPF}, when all input modes are uniformly excited. The mean power fraction of modal groups is represented vs. the group eigenvalues, defined as $\epsilon_j=\beta_j-\beta_{j=Q}$; group $j=1$ has the larger eigenvalue. In the figure, both output distributions including and not including MDL are reported, demonstrating that the preferential power flow toward the fundamental mode is not solely related to modal losses. 

\unskip

%\unskip

It was demonstrated in \cite{Zitelli:10848169, Zitelli:lpor.19.202400714} that any physical process (linear or nonlinear) capable of mixing power among all modes, with negligible variation in the total number of photons $N=\sum_jn_j$ and internal energy $U=\sum_j\beta_jn_j$, being $n_j$ the population of a group, produces an output modal distribution described by a weighted Bose–Einstein function (wBE), which corresponds to a maximum Boltzmann entropy of the multimode system. The mean modal power fraction of group $j$ is

\begin{equation}
\lvert f_j \rvert ^2=\frac{2(g_j-1)}{g_j\gamma}\frac{1}{\exp\big(-\frac{\mu'+\epsilon_j}{T}\big)-1}    .
\label{eq:BE}
\end{equation}

In Eq.~\ref{eq:BE}, $\lvert f_j \rvert^2=2 n_j/(\gamma n_0 g_j)$ is the mean modal power fraction, over two polarizations, in modal group $j$. $\gamma$ is proportional to the input pulse power, $\mu'=\mu+\beta_{j=Q}$, with $\mu$ (1/m) a chemical potential and $T$ (1/m) an optical temperature. 
%In the experiments, $\mu'$, $T$ and $\gamma$ are free parameters at only one intermediate power; for other powers, $\gamma$ is scaled proportionally to $N$ or $P$, and $\mu'$, $T$ are the only fitting parameters. The value of $n_0$ is calculated from $\gamma$ and $N$ at a same power level.
$\mu'$ and $T$ are two degrees of freedom for fitting Eq.~\ref{eq:BE} to the experimental data. The $\gamma$ parameter is free at only one intermediate pulse energy; for other energies, it must scale proportionally to the input power or $N$; the second constraint is the respect of the conservation law $\sum_{j=1}^{Q}(g_j/2)\lvert f_j \rvert ^2=1$.

The assumption of negligible variation for $U$ and $N$, a condition for the validity of Eq.~\ref{eq:BE}, is generally verified in experiments where the RMC is responsible for the power exchange between modes. For example, in the case of transmission in an OM4 GRIN fiber, the modes supported at $\lambda=1550$ nm are $M=55$ per polarization, for a total of $Q=10$ modal groups; the propagation constants of the different groups vary from $\beta_1=5.91\times10^6$ m$^{-1}$ to $\beta_{10}=5.86\times10^6$ m$^{-1}$. Even in the extreme case in which the RMC causes a total transfer of power from group $j=10$ to group $j=1$, the internal energy variation is $\Delta U/U=(\beta_{10}-\beta_1)/\beta_{10}=-0.0087$. The number of photons $N$ in the case of linear transmission is represented by the average power of the signal, or by the energy of the pulses launched into the fiber; the variation of $N$ is negligible as long as the linear losses of the fiber are small, i.e. up to a few kilometers of fiber.

An alternative law, the Rayleigh-Jeans (RJ) \cite{Wu2019} is commonly used in the literature when nonlinear optical effects cause the modal power scrambling; we will see that this second law is not able to properly describe the distribution of the LOMs in the linear case, when RMC is the responsible of modal power flow.

In Fig.~\ref{fig3} it is reported a fit of the output distribution normalized to the MDL, using Eq.~\ref{eq:BE}. Correspondence of the experimental distribution to the wBE is good, with parameters $T=4.73\times10^6$ m$^{-1}$, $\mu'=-6.98\times10^4$ m$^{-1}$, $\gamma=3715$ and with a fit R-squared of 0.97. Hence, a uniform modal excitation at input has provided non-uniform modal distribution at the output, even after removing the modal loss.

\subsection{Linear Regime Simulation}

In order to demonstrate the universality of the steady state distributions observed in the experiment of Fig.~\ref{fig3}, numerical simulations of linear regime transmission were performed in a multimode OM4 fiber, using the coupled-mode GNLSE model of Eqs.~\ref{eq:GNLSE} to \ref{eq:Deltan}.

A realistic transmission system was tested, with modal channels composed by trains of 64 encoded pseudo-random pulses at 40 Gb/s per channel, $\lambda=1550$ nm and femto-Joule pulse energy to prevent the nonlinear effects. $Q=10$ modal groups and $M=55$ modes were coupled at the input with same power; hence, 55 coupled Eqs.~\ref{eq:GNLSE} were used. Transmission distance was $L=1$ km and RMC correlation length $L_c=2$ m; the RMC randomly scrambled the modal power 500 times over the transmission line; a large $RMC_{ext}=10^{-5}$ was set to observe a considerable power flow.
The modal distribution was recorded after each RMC step, repeating the runs 10 times. To reduce statistical fluctuations, the distributions were averaged over 10 runs, and a moving average over 25 consecutive $L_c$ steps (50 m) was then applied.

\begin{figure}[H]
%\isPreprints{\centering}{} % Only used for preprints
\centering
\includegraphics[width=12.0 cm]{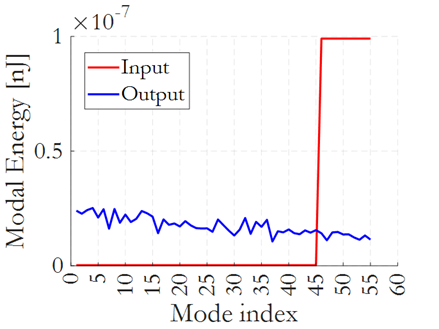}
\caption{Simulated input and output modal energy distributions when input group $10$ is launched.\label{fig4}}
\end{figure}

\begin{figure}[H]
%\isPreprints{\centering}{} % Only used for preprints
\centering
\includegraphics[width=12.0 cm]{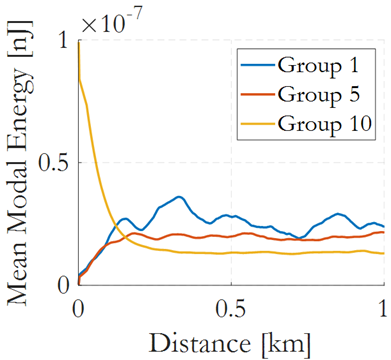}
\caption{Average modal energy in groups $1$, $5$ and $10$ vs. distance, in the simulation of Fig.~\ref{fig4}.\label{fig5}}
\end{figure}

As a first simulation, equal input power was applied only to modes $46$ to $55$ (modal group $10$) leaving the other groups unpopulated. Figure~\ref{fig4} shows the modal energy distribution at the input and output (energy is calculated over the simulation window propagating 32 pulses and 32 spaces per mode); power appears to re-organize to a non-uniform output distribution which promotes the LOMs. The mean energy of the modal groups $1$, $5$ and $10$ is represented in Fig.~\ref{fig5} vs. distance; group $1$ ($10$) experiences the larger modal energy increase (reduction), demonstrating that the RMC-induced power flow promotes the LOMs at the expense of the HOMs ; we recall that the simulation did not include modal losses.

\unskip

\unskip

\begin{figure}[H]
%\isPreprints{\centering}{} % Only used for preprints
\centering
\includegraphics[width=12.0 cm]{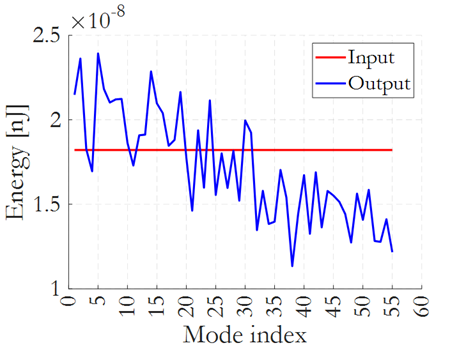}
\caption{Simulated input and output modal energy distributions when input modes are launched with uniform power.\label{fig6}}
\end{figure}   
\unskip

\begin{figure}[H]
%\isPreprints{\centering}{} % Only used for preprints
\centering
\includegraphics[width=12.0 cm]{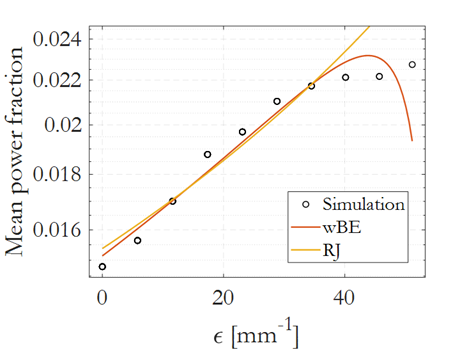}
\caption{Mean modal power fraction vs. modal eigenvalues, for the simulation of Fig.~\ref{fig6}. A wBE and RJ fit is calculated over $10$ modal groups.\label{fig7}}
\end{figure}   

As a second case, equal power was launched over the $55$ modes, obtaining the output energy distribution of Fig.~\ref{fig6}, and the mean modal power fraction of Fig.~\ref{fig7}. Also in this case, RMC caused the modal power to reach a steady state promoting the LOMs. The mean modal power fraction $\lvert f_j \rvert^2$ was properly fitted by a wBE law with $T=7.99\times10^5$ m$^{-1}$, $\mu'=-1.01\times10^5$ m$^{-1}$, $\gamma=941$ and with a fit R-squared of $0.96$. For comparison, the data fit using an RJ was unable to represent properly the LOMs but obtained a good R-squared of $0.93$.

\subsection{Power Flow Simulation}
%\subsubsection{LinSim}

Power-flow numerical model is alternative to the GNLSE and capable of describing the exchange of modal power caused by the RMC; the model treats $P_j$ as the total power propagating in modal group $j$, and does not deal with intra-group modal coupling. The model was derived differently from the GNLSE, and constitutes a valid test bed to demonstrate the formation of steady states characterized by non-uniform modal group distributions, both in the presence and absence of MDL.

\begin{figure}[H]
%\isPreprints{\centering}{} % Only used for preprints
\centering
\includegraphics[width=12.0 cm]{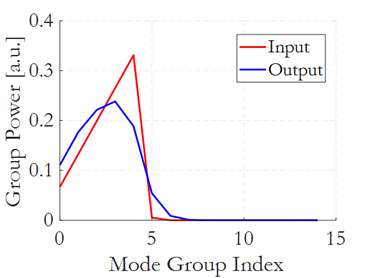}
\caption{Input and output modal group powers simulated by the power-flow model.\label{fig8}}
\end{figure}

\begin{figure}[H]
%\isPreprints{\centering}{} % Only used for preprints
\centering
\includegraphics[width=12.0 cm]{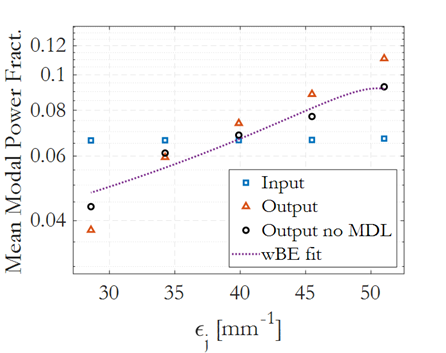}
\caption{Mean modal power fractions vs. eigenvalues, for the simulation with uniform input power, $A=4.4\times10^{-6}$ m$^{-1}$ or $A=0$.\label{fig9}}
\end{figure}

In Fig.~\ref{fig8}, equal modal power was launched in the first $5$ modal groups, and no power into groups $6$ to $15$; the input curve in the figure represents the corresponding total power in the groups. Equation~\ref{eq:PowerFlow} was used for propagation with $D=8.5\times10^{-5}$ m$^{-1}$ and $A=4.4\times10^{-6}$ m$^{-1}$.
Figure shows that the output group power diffuses promoting the LOMS. This is best highlighted in Fig.~\ref{fig9}, where it is reported the mean modal power fraction vs. eigenvalues in the presence and absence of MDL ($A=0$). Also in the absence of MDL, the fundamental mode carries a larger power than the modes of group $5$. A wBE fit properly describes the case 
$A=0$, obtaining $T=1.15\times10^7$ m$^{-1}$, $\mu'=-6.24\times10^4$ m$^{-1}$, $\gamma=12800$ and with a fit R-squared of $0.95$. 

\unskip

\subsection{Quantum Experiment}
%\subsubsection{LinSim}

\begin{figure}[H]
%\isPreprints{\centering}{} % Only used for preprints
\centering
\includegraphics[width=12.0 cm]{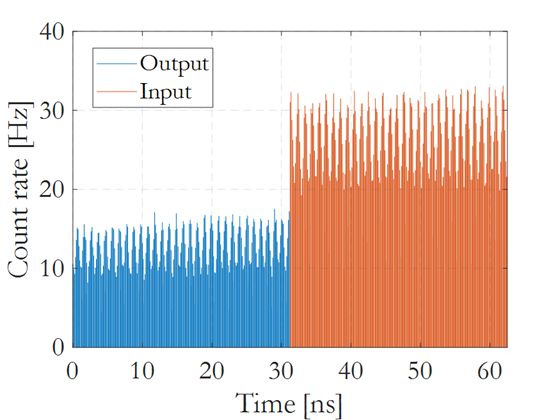}
\caption{Count rate histograms of the single-photon pulses at the input (in orange) and output (in blue) of the modal channels.\label{fig10}}
\end{figure}

A challenging question is whether multimode fiber transmission in the quantum regime, with pulses composed of single photons, exhibits power exchange characteristics between modes similar to those observed in the linear regime. In other words, whether the RMC's property of promoting LOMs has ergodicity characteristics. The experiment of Fig.~\ref{fig1} was repeated after attenuating input pulses to $\mu=1$ photon per pulse per input mode; the same quantum power was applied to the 15 input modes and was constantly monitored at the input. Photon counts were recorded at the modal inputs and outputs, by replacing the power meters with photon counting techniques, as explained in the methods section.
Figure~\ref{fig10} is an example reporting a train of pulses at the modal inputs (in orange color) and outputs (in blue) at a repetition frequency of 1 Gpulse/s.

\begin{figure}[H]
%\isPreprints{\centering}{} % Only used for preprints
\centering
\includegraphics[width=12.0 cm]{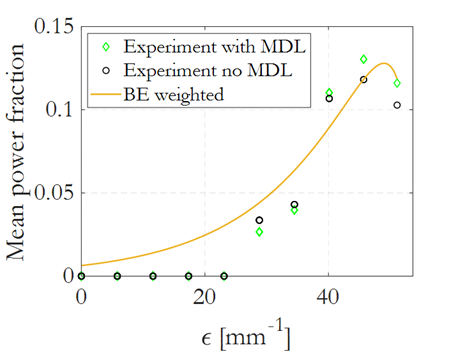}
\caption{Quantum mean modal power fractions $\lvert f_j \rvert^2$ vs. eigenvalues recorded in the first 5 modal groups, with and without MDL. The wBE fit is calculated from the lossless distribution.\label{fig11}}
\end{figure}

Figure~\ref{fig11} reports the mean (quantum) modal power fractions $\lvert f_j \rvert^2$ vs. eigenvalues recorded in the first 5 modal groups, supposing the fiber supports 10 groups. Powers were obtained from output modal counts, after removing or not removing the MDL using Eq.~\ref{eq:MMPF}. The obtained quantum distribution effectively recalls the ones observed in the linear regime experiment of Fig.~\ref{fig3} and in the simulation of Fig.~\ref{fig7}. The wBE law was able to fit the quantum distribution obtaining $T=1.44\times10^4$ m$^{-1}$, $\mu'=-1.48\times10^5$ m$^{-1}$, $\gamma=0.01$ and a fit R-squared of $0.97$.

\unskip

\section{Discussion and Conclusions}

All the results presented in this work, the experiment in the linear regime, numerical simulations using the coupled GNLEs, the power-flow simulations and the experiment in the quantum regime, confirm that a sufficiently long multimode fiber produces steady state nonuniform modal distributions, even when equal power is injected in all input modes. The MDL enhances the effect but is not the only responsible. 

The length of fiber required to produce a steady state was estimated in \cite{SAVOVIC2018223}, ranging from a few hundred meters to a few kilometers depending on the magnitude of the parameter $D$. Consequently, it is not possible to effectively study the effects of RMC using a few meters of fiber, and it is instead necessary to precisely eliminate the effects of MDL from the experimental results.

When in steady state, the output distributions produced by the RMC alone are characterized by higher power in the LOMs and are properly fitted by the weighted Bose-Einstein law \cite{Zitelli:lpor.19.202400714,Zitelli:10848169} with fit R-squared greater than 0.95. The distributions are confirmed by the experiments in quantum regime, confirming that the RMC has ergodic properties: it is possible to obtain similar distributions measuring the optical power from multi-photon pulses or counting repeated single-photon pulses. This property paves the wave to quantum thermodynamics experiments that reconcile the thermodynamics of optical multimodal systems with quantum optics.

\section{Funding}

Project ECS 0000024 Rome Technopole, Funded by the European Union – NextGenerationEU

%\bibliographystyle{unsrt}
%\bibliography{References.bib,biblio.bib}

%%%%%%%%%% If preparing manually:

\end{document}